# Efficient Approaches for Designing Fault Tolerant Reversible Carry Look-Ahead and Carry-Skip Adders

Md. Saiful Islam, Muhammad Mahbubur Rahman, Zerina begum, and Mohd. Zulfiquar Hafiz

*Abstract*—Combinational or Classical logic circuits dissipate heat for every bit of information that is lost. Information is lost when the input vector cannot be recovered from its corresponding output vector. Reversible logic circuit implements only the functions having one-to-one mapping between its input and output vectors and therefore naturally takes care of heating. Reversible logic design becomes one of the promising research directions in low power dissipating circuit design in the past few years and has found its application in low power CMOS design, digital signal processing and nanotechnology. This paper presents the efficient approaches for designing fault tolerant reversible fast adders that implement carry look-ahead and carry-skip logic. The proposed high speed reversible adders include MIG gates for the realization of its basic building block. The MIG gate is universal and parity preserving. It allows any fault that affects no more than a single signal readily detectable at the circuit's primary outputs. It has also been demonstrated that the proposed design offers less hardware complexity and is efficient in terms of gate count, garbage outputs and constant inputs than the existing counterparts.

## I. INTRODUCTION

IRREVERSIBLE circuits dissipate heat in the amount of $kT \ln 2$ Joule for every bit of information that is lost irrespective of their implementation technologies, where $k$ is the Boltzmann constant and $T$ is the operating temperature [1]. Information is lost when the circuit implements nonbijective functions. Therefore in irreversible logic circuit the input vector cannot be recovered from its output vectors. Reversible logic circuit by definition realizes only those functions having one-to-one mapping between its input and output assignments. Hence in reversible circuits no information is lost. According to [2] zero energy dissipation would be possible only if the network consists of reversible gates. Thus reversibility will become an essential property in future circuit design.

Reversible logic imposes many design constraints that need to be either ensured or optimized for implementing any particular Boolean functions. Firstly, in reversible logic circuit the number of inputs must be equal to the number of outputs. Secondly, for each input pattern there must be a unique output pattern. Thirdly, each output will be used only once, that is, no fan out is allowed. Finally, the resulting circuit must be acyclic [3]-[5]. Any reversible logic design should minimize the followings [6]:

- **Garbages**: outputs that are not used as primary outputs are termed as garbages
- **Constants**: constants are the input lines that are either set to zero(0) or one (1) in the circuit's input side
- **Gate Count**: number of gates used to realize the system
- **Hardware Complexity**: refers to the number of basic gates (NOT, AND and EXOR gate) used to synthesize the given function

Parity checking is one of the widely used mechanisms for detecting single level fault in communication and many other systems. It is believed that if the parity of the input data is maintained throughout the computation, no intermediate checking would be required [7], [8]. Therefore, parity preserving reversible circuits will be the future design trends towards the development of fault tolerant reversible systems in nanotechnology. And a gating network will be parity preserving if its individual gate is parity preserving [7]. Thus, we need parity preserving reversible logic gates to construct parity preserving reversible circuits. This paper presents the efficient approaches for designing fault tolerant reversible carry look-ahead and carry-skip adders. To design the basic building block of both carry look-ahead adder (CLA) and carry-skip adder (CSA) we have used IG gate proposed in [9], [10]. The IG gate is parity preserving and complete. Finally, this paper presents a 16-bit high speed fault tolerant reversible adder that is efficient in terms of gate count, garbage outputs, constant inputs and hardware complexity.

The rest of the paper is organized as follows: section II presents reversible logic and some basic reversible logic gates, section III presents parity preserving reversible logic gates, section IV presents the efficient approaches for designing CLA and CSA, section V evaluates proposed designs in terms of gate count, hardware complexity,

Md. Saiful Islam is with the Institute of Information Technology, University of Dhaka, Dhaka-1000, Bangladesh (corresponding author, phone: +880-1911489986; fax: +880-2-8615583; e-mail: saifulit@ univdhaka.edu).

Muhammad Mahbubur Rahman is with the Dept. of Computer Science, American International University- Bangladesh, Dhaka-1213, Bangladesh (e-mail: mus_mahbub@hotmail.com).

Zerina Begum is with the Institute of Information Technology, University of Dhaka, Dhaka-1000, Bangladesh (e-mail: zerin@univdhaka.edu).

Mohd. Zulfiquar Hafiz is the Director of Institute of Information Technology, University of Dhaka, Dhaka-1000, Bangladesh (e-mail: jewel@univdhaka.edu).

garbages and constant inputs; and finally section VI concludes the paper.

## II. REVERSIBLE LOGIC GATES

A gate that implements any bijective function involving $n$ inputs and $n$ outputs is called an $n*n$ reversible logic gate. There exist many reversible gates in the literature. Among them 2*2 Feynman gate [11] (shown in Fig. 1), 3*3 Fredkin gate [12] (shown in Fig. 2), 3*3 Toffoli gate [13] (shown in Fig. 3) and 3*3 Peres gate [14] (shown in Fig. 4) are the most referred. Feynman (FG), Fredkin (FRG) and Peres (PG) gates are one through gates, that is, one of its output lines is identical to one of its input lines. On the other hand, Toffoli gate is two through, that is, two of its outputs are identical to two of its inputs. It can be easily verified that all of these gates are reversible. Each gate has an equal number of input and output lines. For each input combination there is a unique output combination.

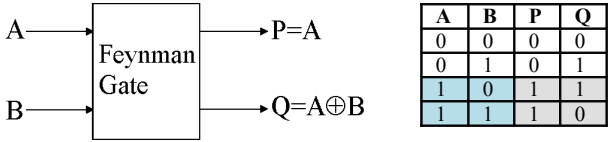

Fig. 1. 2*2 Feynman gate.

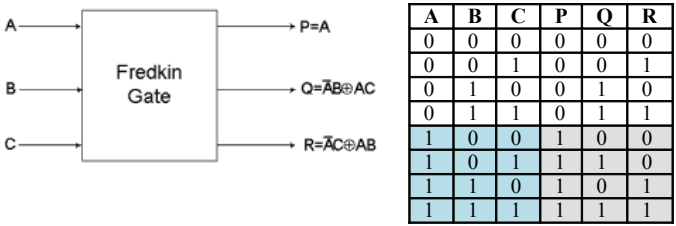

Fig. 2. 3*3 Fredkin gate

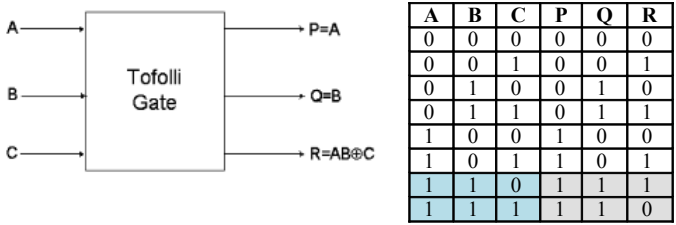

Fig. 3. 3*3 Toffoli gate.

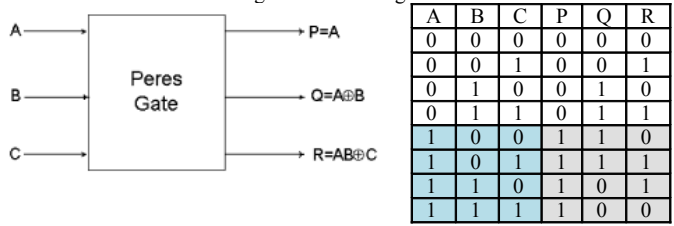

Fig. 4. 3*3 Peres gate.

### III. PARITY PRESERVING REVERSIBLE GATES

A reversible gate will be parity preserving if the parity of the inputs matches the parity of the outputs. Mathematically, a reversible gate having $n$ input lines and $n$ output lines will be parity preserving if and only if:

$$I_1 \oplus I_2 \oplus \cdots \oplus I_n \leftrightarrow O_1 \oplus O_2 \oplus \cdots \oplus O_n$$

where $I_i$ and $O_j$ are the input and output lines. Not all of the gates presented in section II are parity preserving. Only Fredkin gate is parity preserving and it can be easily verified by examining its truth table. That is, Fredkin gate maintains $A \oplus B \oplus C \leftrightarrow P \oplus Q \oplus R$. It can also be said that the Fredkin gate is zero preserving (once preserving as well) and therefore conservative [15]. Other parity preserving reversible logic gates are 3*3 Feynman Double gate [7] (shown in Fig. 5), 3*3 New Fault Tolerant gate [16] (shown in Fig. 6) and newly proposed 4*4 IG gate [9], [10] (shown in Fig. 7). Feynman Double gate can be as used as the fault tolerant copying gate when it's B and C input lines are set to constants.

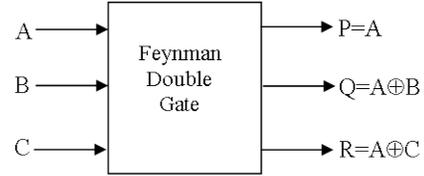

Fig. 5. 3*3 Feynman Double gate.

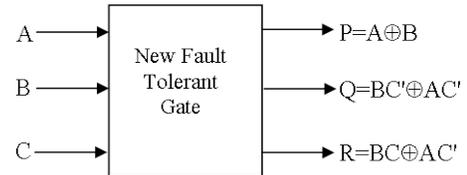

Fig. 6. 3*3 New Fault Tolerant gate

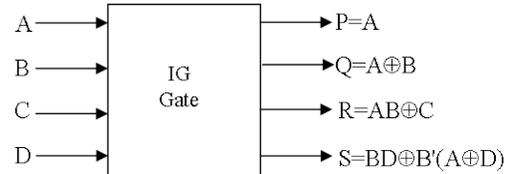

Fig. 7. 4*4 IG gate

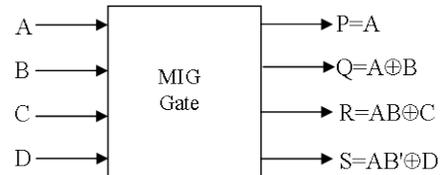

Fig. 8. 4*4 MIG gate

The first three output lines of IG gate produce the same output as PG gate. The fourth one can be considered as garbage if we wish to replace PG by IG. It is mainly introduced for preserving the parity. The fourth output line of IG gate can also be minimized to reduce the hardware complexity as follows:

$$BD \oplus B'(A \oplus D) \Leftrightarrow AB' \oplus D$$

The modified version of IG is depicted in Fig. 8.

### IV. SYNTHESIS OF FAULT TOLERANT REVERSIBLE CLA AND CSA

The basic building block of many complex computational systems is the full adder (FA). Both CLA and CSA include full adders. Realization of the efficient reversible full adder circuit given in [3]-[5] includes two 3*3 Peres gates. The circuit is minimized in terms of gate count, garbage outputs, constant inputs and hardware complexity. It has also been proved in [3]-[5] that a reversible full adder circuit can be realized with at least two garbage outputs and one constant input.

Fault tolerant logic synthesis of reversible full adder circuit requires that its individual gate unit must be fault tolerant reversible gates. It has been proved in [9], [10] that a fault tolerant reversible full adder circuit requires at least three garbage outputs and two constant inputs. The fault tolerant reversible full adder circuit presented in [9], [10] is given below:

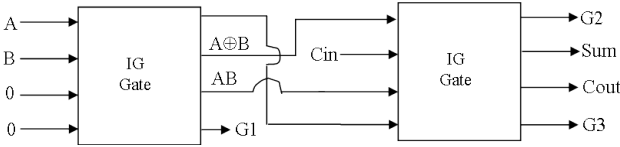
Fig. 9. Fault tolerant reversible full adder circuit includes two 4*4 IG gates [9], [10].

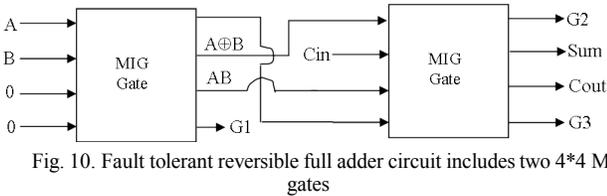
Fig. 10. Fault tolerant reversible full adder circuit includes two 4*4 MIG gates

This paper will use MIG gate for fault tolerant reversible full adder implementation and will thereby minimize the hardware complexity of the presented system (shown in Fig. 10). The block diagram of the fault tolerant reversible full adder can be depicted as follows:

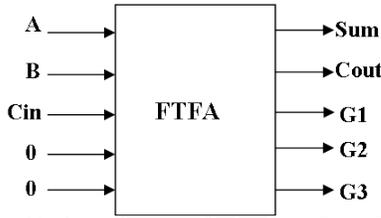
Fig. 11. Block diagram of fault tolerant full adder.

### A. Ripple Carry Adder

The most straightforward realization of a final stage adder for two $n$-bit operands is ripple carry adder. The RCA requires $n$ full adders (FAs). The carry out of the $i^{th}$ FA is connected to the carry in of the $(i+1)^{th}$ FA. The Fig. 12 shows a reversible logic implementation of an $n$-bit final stage fault tolerant ripple carry adder.

### B. Carry Look-Ahead Adder

The main idea behind carry look-ahead addition is an attempt to generate all incoming carries in parallel and avoid waiting until the correct carry propagates from the stage (FA) of the adder where it has been generated. The Fig. 13 shows a reversible logic implementation of a 2-bit fault tolerant carry look-ahead adder.

### C. Carry-Skip Adder

A carry-skip adder reduces the carry-propagation time by skipping over groups of consecutive adder stages. The carry-skip adder is usually comparable in speed to the carry look-ahead technique, but it requires less chip area and consumes less power. The Fig. 14 shows a reversible logic implementation of a 4-bit fault tolerant carry skip adder.

### D. A Proposal for a Novel 16-bit Fault Tolerant Reversible Adder

This paper proposes a 16-bit high speed adder that includes four fixed-size blocks, each of size 4, shown in Fig. 15. Each block is a 4-bit *ripple carry adder* and block *carry in* is skipped using *carry skip logic*. Therefore, each block can be operated in parallel and will thereby reduce the overall delay.

## V. EVALUATION OF THE PROPOSED DESIGNS

The presented adder circuits can be evaluated in terms of hardware complexity, gate count, constant inputs and garbage outputs produced. Evaluation of the proposed circuit can be comprehended easily with the help of the comparative results given in Table I.

### A. Hardware Complexity

One of the main factors of a circuit is its hardware complexity. It can be proved that the proposed adder circuits are better than the existing approaches in terms of hardware complexity. Let

$\alpha$ = A two input EXOR gate calculation
$\beta$ = A two input AND gate calculation
$\delta$ = A NOT gate calculation
T = Total logical calculation

The FTFA and RCA given in [9], [10] have total logical calculation T=$8\alpha+6\beta+2\delta$ and T=$32\alpha+24\beta+8\delta$ respectively. The FTFA given in [15] has total logical calculation T=$8\alpha+16\beta+4\delta$. The presented FTFA and RCA have total logical calculation T=$6\alpha+4\beta+2\delta$ and T=$24\alpha+16\beta+8\delta$ respectively. Therefore, we can say that the presented FTFA and RCA is better than the design given in [9], [10] and [15] in terms of hardware complexity. The CSA with fanout given in [15] has total logical calculation T=$40\alpha+80\beta+20\delta$. The presented CSA without fanout has total logical calculation T=$40\alpha+28\beta+12\delta$. The total logical calculation of the presented CLA and HSA are T=$47\alpha+23\beta+9\delta$ and T=$320\alpha+112\beta+48\delta$ respectively.

### B. Gate Count

The presented FTFA and RCA, and the design given in [9], [10] require the same number of reversible gates. The FTFA given in [15] requires 4 reversible gates. Hence the presented FTFA is better than the design given in [15] in terms of gate count. The presented CSA requires 14 reversible gates whereas the CSA given in [15] requires 20 reversible gates. Therefore, we can say that the presented CSA is better than the design given in [15] in terms of gate count. The proposed CLA and HSA require 19 and 56 reversible gates respectively.

### C. Garbage Outputs

Garbage output refers to the output of the reversible gate that is not used as a primary output or as input to other gates. One of the other major constraints in designing a reversible logic circuit is to lessen number of garbage outputs [5]. The presented FTFA and RCA, and the design given in [9], [10] and [15] produce same number garbage outputs. The presented CSA produces 19 garbage outputs whereas the design given in [15] produces 16 garbage outputs only. This is because the presented designs do not allow fanout. Fanout is strictly prohibited in reversible logic design. The proposed CLA and HSA produce 28 and 76 garbage outputs respectively.

*D. Constant Inputs*

Number of constant inputs is one of the other main factors in designing a reversible logic circuit. The input that is added to an $n \times k$ function to make it reversible is called constant input [5]. The presented FTFA and RCA, and the design given in [9], [10] and [15] require the same number constant inputs. The presented CSA requires 15 constant inputs whereas the design given in [15] produces 11 constant inputs only. This is because the presented designs do not allow fanout. Fanout is strictly prohibited in reversible logic design. The proposed CLA and HSA require 26 and 60 constant inputs respectively.

VI. CONCLUSION

This paper presents the efficient approaches for designing fault tolerant reversible carry look-ahead and carry-skip adders. The proposed designs are optimized in terms of gate count, garbage outputs, constant inputs and hardware complexity. Finally a novel design of 16-bit high speed fault tolerant reversible adder is proposed. It has also been demonstrated that the proposed designs offer less hardware complexity and are efficient in terms of gate count, garbage outputs and constant inputs.

**Md. Saiful Islam** joined the Institute of Information Technology at the University of Dhaka in January 2008. He holds B. Sc (Hons) and M.S. degree in Computer Science and Engineering from the University of Dhaka. He has more than 25 research articles published in several journals and conferences. He also leads and teaches modules at both B. Sc and MSc levels in Information Technology and Software Engineering. His research interests include Reversible Logic Design, Reconfigurable Computing, Multimedia Information Retrieval, Machine Learning, Evolutionary Computing and Bioinformatics.

**Muhammad Mahbubur Rahman** received his master degree in Information Technology from Institute of Information Technology at University of Dhaka. Currently, he is working as a lecturer in the Department of Computer Science at American International University-Bangladesh. His research interests include Data Mining, Machine Learning, Bioinformatics, Game Theory, Artificial Intelligence, Econometrics and Logic Circuit Minimization.

**Dr. Zerina Begum** obtained her B. Sc (Hons) and M. Sc in Physics from the University of Dhaka in1986 and 1987 respectively. She started her carrier as a Research Fellow in Bose Research Centre at University of Dhaka. Two years later she joined as a Programmer in the Computer Centre of the same university. She was awarded M. Phil for her research in Semiconductor Technology and Computerized Data Acquisition System in 1995. As recognition of her outstanding research in organic semiconductor, Department of Physics, University of Dhaka awarded her with Ph. D. degree in 2007. She is still continuing to serve University of Dhaka as a Professor in the Institute of Information Technology. Her current research interest areas include Semiconductor Technology, Computerized Data Acquisition System, Advanced Digital Electronics and Software Engineering.

**Mohd. Zulfiquar Hafiz** obtained his B. Sc (Hons) and M. Sc in Pure Mathematics from the University of Dhaka in 1986 and 1987 respectively. He started his carrier as a programmer in the Computer Centre at University of Dhaka in 1993. He was repositioned as a Lecturer in the centre in 1997. He is still continuing to serve University of Dhaka as an Associate Professor in the Institute of Information Technology. His current research interest areas include Computer Networks, Embedded Systems and Computational Fluid Dynamics.

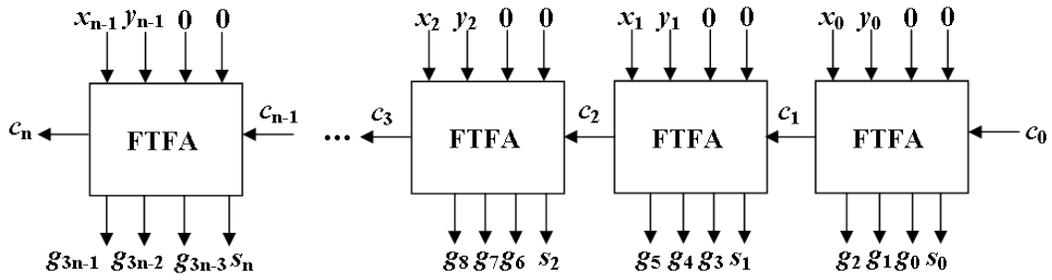

Fig. 12. Reversible logic implementation of fault tolerant *n*-bit ripple carry adder.

TABLE I  COMPARATIVE EXPERIMENTAL RESULT OF DIFFERENT FAULT TOLERANT REVERSIBLE ADDER CIRCUITS

| Design | Gate Count | Hardware Complexity | Constant Inputs | Garbage Outputs |
|---|---|---|---|---|
| 1-bit FTFA | 2 MIG | $6\alpha+4\beta+2\delta$ | 2 | 3 |
| 1-bit FTFA [9][10] | 2 IG | $8\alpha+6\beta+2\delta$ | 2 | 3 |
| 1-bit FTFA [15] | 4 FRG | $8\alpha+16\beta+4\delta$ | 2 | 3 |
| 4-bit RCA | 8 MIG | $24\alpha+16\beta+8\delta$ | 8 | 12 |
| 4-bit RCA [9][10] | 8 IG | $32\alpha+24\beta+8\delta$ | 8 | 12 |
| 4-bit RCA [15] | 16 FRG | $32\alpha+64\beta+16\delta$ | 8 | 12 |
| 2-bit CLA | 4 MIG+10 F2G+ 5 NFT = 19 | $47\alpha+23\beta+9\delta$ | 26 | 28 |
| 4-bit CSA | 8 MIG + 4NFT +2 F2G=14 | $40\alpha+28\beta+12\delta$ | 15 | 19 |
| 4-bit CSA [15] | 20 FRG | $40\alpha+80\beta+20\delta$ | 11 | 16 |
| 16-bit HSA | 32 MIG+ 16NFT+8F2G = 56 | $320\alpha+112\beta+48\delta$ | 60 | 76 |

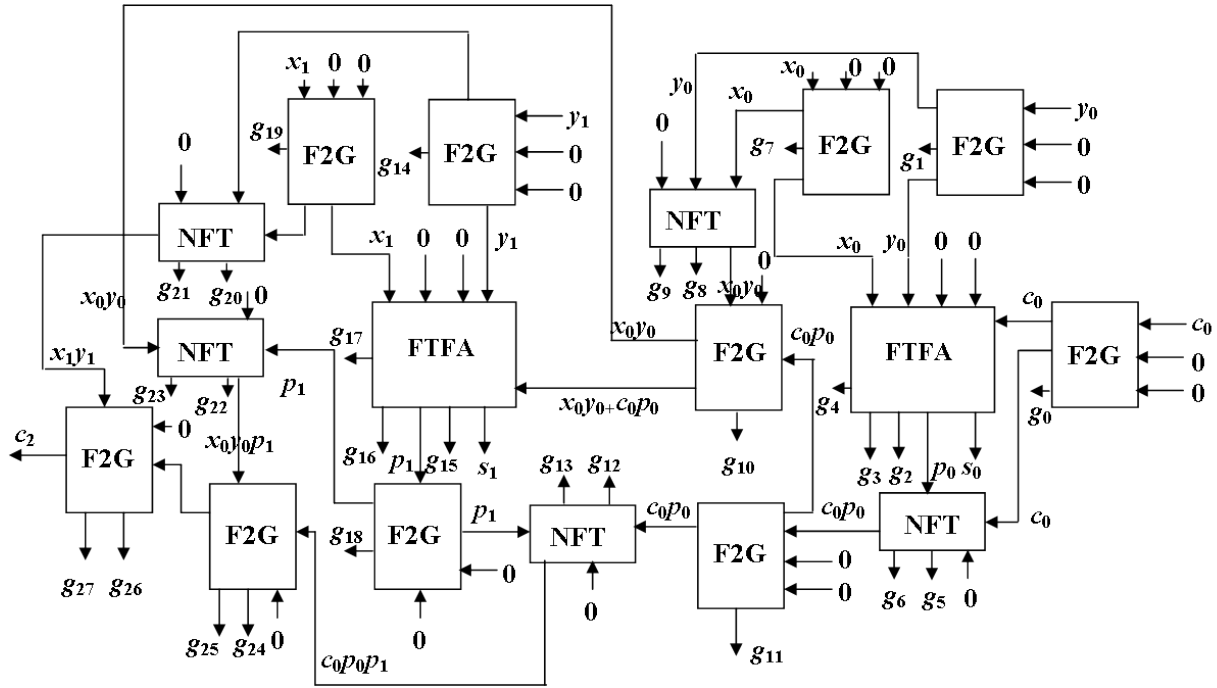

Fig. 13. Reversible logic implementation of 2-bit fault tolerant carry look-ahead adder.

Fig. 13 presents a reversible logic implementation of a 2-bit fault tolerant carry look-ahead adder. The realization requires 19 parity preserving reversible gates of which are 5 NFTs, 10 F2Gs and 8 MIGs. It needs 26 constant inputs and produces 28 garbage outputs.

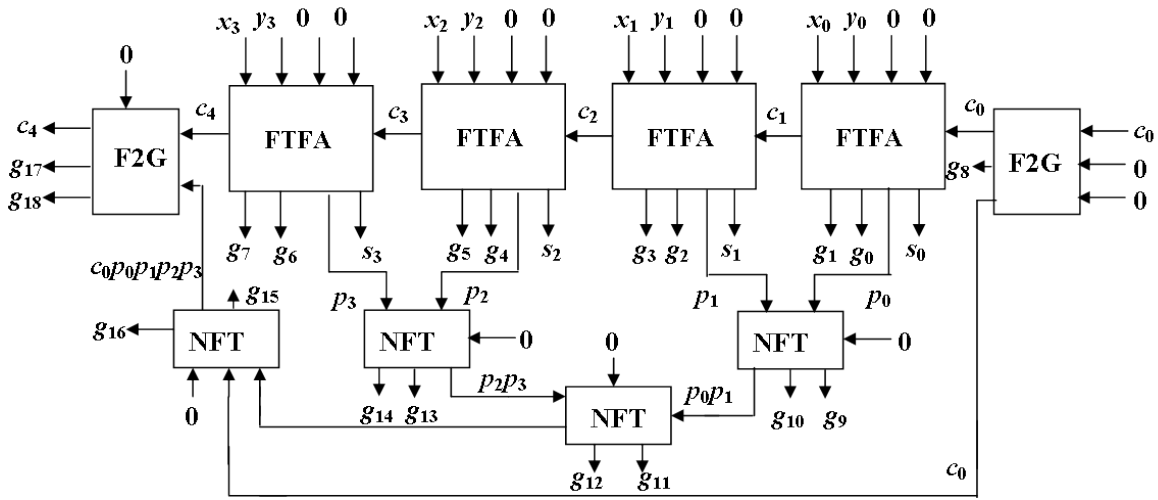

Fig. 14. Reversible logic implementation of a 4-bit fault tolerant carry skip adder.

Fig. 14 presents an efficient reversible logic implementation of a 4-bit fault tolerant carry skip adder. The realization requires 14 parity preserving reversible gates of which are 4 NFTs, 2 F2Gs and 8 MIGs. It needs 15 constant inputs and produces 19 garbage outputs. The proposed CSA is better than the CSA presented in [15] in terms of gate count and hardware complexity. The proposed design requires more garbages than the design given in [15]. This is because fanout are not allowed in the proposed design since it is strictly prohibited in reversible logic design. The 16-bit high speed fault tolerant reversible adder is realized by dividing it into four fixed-sized blocks. Each block implements *carry skip logic* to skip block *carry in* for the next block. To realize the block 4-bit RCA is proposed instead of 4-bit CLA because CLA is more costly than CSA in terms of gate count. The fault tolerant CSA offers less hardware complexity than CLA too.

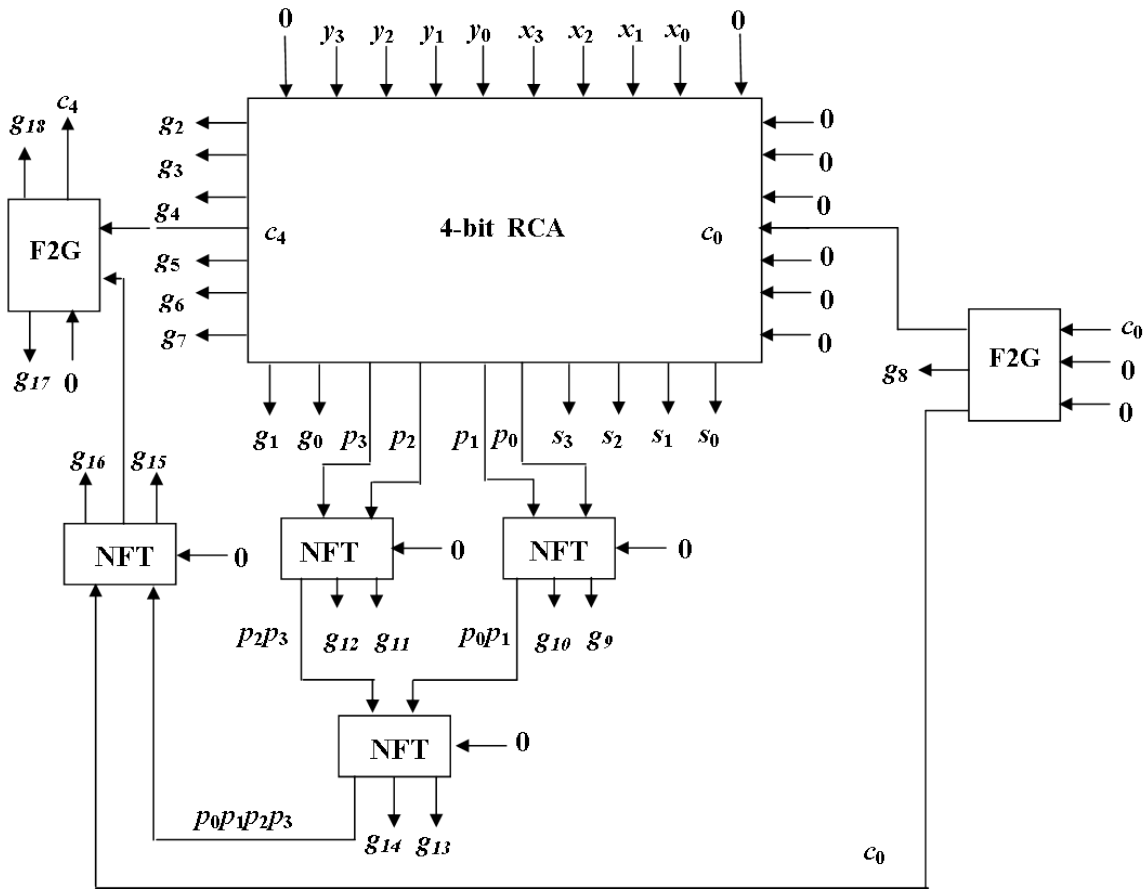

Fig. 15. Block diagram of a 4-bit carry skip adder.